\documentclass[preprint2]{aastex}

\shorttitle{Hypervelocity stars}
\shortauthors{Abadi, Navarro \& Steinmetz}

\begin{document}

\title{An Alternative Origin for Hypervelocity Stars}

\author{Mario G. Abadi}
\affil{Instituto de Astronom\'ia Te\'orica y Experimental (IATE) 
\\Observatorio Astron\'omico de C\'ordoba and CONICET 
\\Laprida 854 X5000BGR C\'ordoba, Argentina
}
\email{mario@oac.uncor.edu}

\author{Julio F. Navarro}
\affil{Department of Physics and Astronomy, University of Victoria 
\\3800 Finnerty Road, Victoria, BC V8P 5C2, Canada}
\email{jfn@uvic.ca}

\and

\author{Matthias Steinmetz}
\affil{Astrophysikalisches Institut Potsdam
\\An der Sternwarte 16, 14482 Potsdam, Germany}
\email{msteinmetz@aip.de}


\begin{abstract}
Halo stars with unusually high radial velocity ("hypervelocity" stars,
or HVS) are thought to be stars unbound to the Milky Way that
originate from the gravitational interaction of stellar systems with
the supermassive black hole at the Galactic center. We examine the
latest HVS compilation and find peculiarities that are unexpected in
this black hole-ejection scenario. For example, a large fraction of
HVS cluster around the constellation of Leo and share a common travel
time of $\sim 100$-$200$ Myr. Furthermore, their velocities are not
really extreme if, as suggested by recent galaxy formation models, the
Milky Way is embedded within a $2.5\times 10^{12} \, h^{-1} \,
M_{\odot}$ dark halo with virial velocity of $\sim 220$ km/s. In this
case, the escape velocity at $\sim 50$ kpc would be $\sim 600$ km/s
and very few HVS would be truly unbound. We use numerical simulations
to show that disrupting dwarf galaxies may contribute halo stars with
velocities up to and sometimes exceeding the nominal escape speed of
the system. These stars are arranged in a thinly-collimated outgoing
``tidal tail'' stripped from the dwarf during its latest pericentric
passage. We speculate that some HVS may therefore be tidal debris from
a dwarf recently disrupted near the center of the Galaxy. In this
interpretation, the angular clustering of HVS results because from our
perspective the tail is seen nearly ``end on'', whereas the common
travel time simply reflects the fact that these stars were stripped
simultaneously from the dwarf during a single pericentric
passage. This proposal is eminently falsifiable, since it makes a
number of predictions that are distinct from the black-hole ejection
mechanism and that should be testable with improved HVS datasets.
\end{abstract}



\section{Introduction}
The existence of ultra high-speed stars in the halo of the Milky Way
was recognized by Hills (1998) as an inevitable consequence of the
presence of a supermassive black hole (SMBH) in a region of high
stellar density such as the Galactic center. Hills estimated that the
disruption of tightly bound binaries in the vicinity of the black hole
may propel stars to speeds exceeding $1000$ km/s, well beyond the
escape velocity from the Milky Way. The discovery of a population of
``hypervelocity'' stars (HVS, for short, Brown et al. 2005) in
radial velocity surveys of faint B-type stars in the Galactic halo
(Brown et al. 2006, 2007a,b) gave strong support to Hills' proposal
and led to substantial theoretical interest in the topic. The latest
compilation by Brown, Geller \& Kenyon (2008) lists 16 HVS, all
receding at speeds exceeding $\sim 300$ km/s and reaching up to $\sim
720$ km/s in the Galactic rest frame.

HVS are widely thought to owe their extreme velocities to Hills'
mechanism, since many of the stars appear to be short-lived, massive
main sequence stars such as those often found near the Galactic
center. Other mechanisms also capable of accelerating stars to high
speeds, such as binary disruption in stellar clusters, have usually
been disfavored on the grounds that the maximum ejection velocity is
unlikely to exceed the escape velocity at the surface of a star, and
thus would only be able to accelerate stars up to no more than a few
hundred km/s. The mediation of the much deeper potential well of a
supermassive black hole thus appears needed to explain the extreme
velocities observed for the HVS population.

Although some HVS have very likely been ejected by the Galactic SMBH,
there are a number of issues that remain unresolved in this
scenario. For example, the velocity distribution of HVS and, in
particular, the lack of {\it very} high-velocity stars (i.e., with
speeds exceeding $1000$ km/s), is not easily understood (see, e.g.,
Sesana, Haardt \& Madau 2007). Nor is the existence of at least one
HVS moving at a speed greater than $600$ km/s in the Galactic rest
frame (HD 271791; Heber et al 2008) whose proper motion apparently
rules out a Galactic center origin.

Although small number statistics may explain away these concerns, we
examine in this Letter further peculiarities in the spatial
distribution and kinematics of HVS and argue that these suggest, at
least for some HVS, an origin different from the SMBH-ejection
mechanism. We begin by examining the distribution of travel times and
the angular clustering of HVS compiled from the literature, and use
numerical simulations to explore an alternative mechanism able to
populate the Galactic halo with high-speed stars. We end by
considering several observational tests that may, in the future, be
used to discriminate between these competing scenarios.

\begin{figure}
\epsscale{.89}
\begin{center}
\plotone{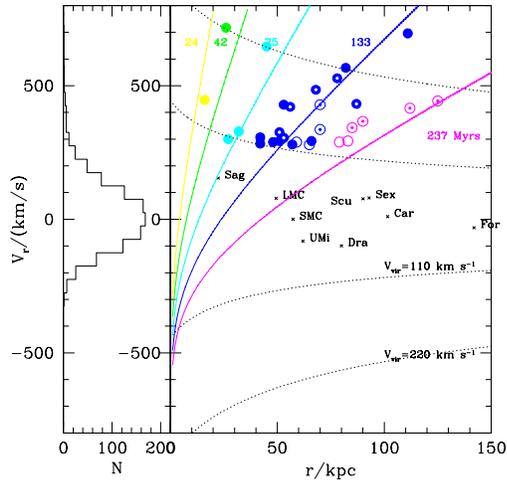}
\end{center}
\caption{
Galactocentric rest frame radial velocity versus distance for all
high-velocity stars with estimated distances compiled from the BGKK
(filled symbols) and BGK08 (open symbols) references.  Symbols with
central dots correspond to stars that fall in the Leo I-centered
circle shown in Fig.~\ref{figs:brown3}. Solid lines show travel times
from the Galactic center in Myrs, as indicated by labels, computed
using the Galactic potential of Bromley et al. (2006). Dotted curves
show the escape velocity of an NFW (Navarro, Frenk \& White 1996,
1997) halo with virial velocity $V_{200}=220$ km/s (the circular velocity
at the solar circle) and $110$ km/s (half of that value), for
reference.  Note that for $V_{200}=220$ km/s only {\it one} HVS is
clearly unbound to the Galaxy.  The left panel shows the radial
velocity histogram for all stars in the BGKK dataset.  Some Milky Way
satellites are also included for comparison.
\label{figs:vr_r}}
\end{figure}

\begin{figure}
\epsscale{.86}
\begin{center}
\plotone{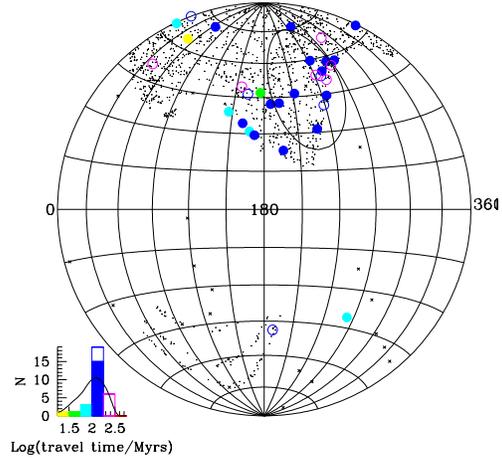}
\end{center}
\caption{ Aitoff projection in Galactic coordinates for all stars in
  the BGKK survey. Stars highlighted in color are high-speed ($V_r \ga
  300$ km/s) stars with estimated distances (filled circles), and
  include a few new stars reported by BGK08 (open symbols). The color
  encodes the travel time of each star, as shown by the histogram in
  the bottom left inset. The filled histogram corresponds to BGKK
  stars, the open histogram adds the new stars in the BGK08
  sample. The solid curve shows the travel-time distribution predicted
  by a model where stars are continuously ejected by the SMBH at the
  Galactic center (Bromley et al. 2006), after taking into account the
  finite lifetime of the stars and the effective survey volume. The
  black circle in the aitoff panel delineates a $26^{\circ}$-diameter
  region in the constellation of Leo. More than half of all HVS with
  $V_r>350$ km/s are contained within that region. This represents a
  significant overdensity of HVS in the direction of Leo, since the
  same region contains only about $\sim 20\%$ of all stars in the
  survey.  A few other Milky Way satellites are shown for reference.
\label{figs:brown3}}
\end{figure}

\section{Analysis}

\subsection{Observational data}

We use primarily data from the Brown et al (2006, 2007a,b, hereafter
BGKK, for short) MMT and Whipple surveys of B-type halo stars selected
from the Sloan Digital Sky Survey.  Our analysis uses Galactic
rest-frame radial velocities, positions in the sky, and distances, as
given by the authors. We note that distance estimates are only listed
for stars with velocities exceeding $\sim 280$ km/s. These estimates
typically assume that the stars are in the main sequence, although it
is difficult to rule out that some (or many) are evolved stars in the
blue horizontal branch (BHB). This is one of the main systematic
uncertainties in our analysis, but it is a shortcoming shared with all
other attempts to study the kinematics of HVS stars with current
datasets.

\subsection{Travel times}

The left panel in Figure~\ref{figs:vr_r} shows the Galactic rest frame
radial velocity distribution of all stars in the BGKK survey, whereas
the right panel shows the radial velocity of all stars with published
distance estimates. These are all high-speed stars ($V_r \ga 280$
km/s) out in the tail of the radial velocity distribution (filled
symbols). Open symbols correspond to the $10$ new stars recently
reported by Brown, Geller \& Kenyon (2008, hereafter BGK08) and are
added here only for completeness, since data for the underlying survey
from which these stars have been identified are not yet publicly
available. (``Dotted'' symbols identify stars in the circle shown in
Fig.~\ref{figs:brown3}, as discussed below.) For illustration, we have
added in Figure~\ref{figs:vr_r} the position in the $r$-$V_r$ plane of
several Milky Way satellite companions, as labeled.

For each star with published distance estimate and radial velocity we
compute a travel time from the Galactic center, assuming purely radial
orbits in the simple Galactic potential model of Bromley et al
(2006). This is given by a spherically symmetric density profile,
$\rho(r)=\rho_0/(1+(r/r_c)^2)$, where $\rho_0=1.396 \times 10^4 \,
M_{\odot} $pc$^{-3}$ and $r_c=8$ pc (Kenyon et al 2008).  The thick
solid curves in Fig.~\ref{figs:vr_r} represent the loci of stars with
constant travel time (labeled in Myr), whereas the dotted lines show
the escape velocity from the Milky Way, assuming it is embedded within
a cold dark matter halo with virial velocity
\footnote{
We assume a Navarro, Frenk \& White (1996, 1997) mass profile for the
halo (with concentration $c=15$) and define the virial velocity as
the circular velocity within a radius, $r_{200}$, enclosing a mean
overdensity of $200$ times the critical value for closure. The virial
velocity defines implicitly the halo mass: for $V_{200}=220$ km/s, the
mass within $r_{200}=220$ kpc/h, is $M_{200}=2.5\times 10^{12}\,
M_{\odot}$/h. We assume h$=0.73$ throughout the paper.}
$V_{200}=220$ km/s and $110$ km/s, respectively. The HVS data in
Figure~\ref{figs:vr_r} is colored according to travel time, binned in
equally spaced logarithmic intervals (see also the histogram in the
inset of Fig.~\ref{figs:brown3}).

Figure~\ref{figs:vr_r} illustrates a few interesting points. One is
that very few stars are actually unbound if the virial velocity of the
Milky Way is of the order of $220$ km/s, the circular speed at the
solar circle. This is actually what is required by recent models of
galaxy formation in order to match simultaneously the zero-point of
the Tully-Fisher relation and the normalization of the galaxy
luminosity function (see, e.g., Croton et al 2006 for a full
discussion). In other words, for $V_{200}=220$ km/s, the HVS
speeds are unusually high, but not necessarily extreme, and only one
HVS would be clearly unbound. This is important, since it suggests
that more prosaic dynamical effects that do not rely on SMBH ejection
may be responsible for the unusual speeds of the HVS population.

The second point to note is that the distribution of HVS travel times
is not uniform. This is illustrated in the histogram shown in the
inset of Fig.~\ref{figs:brown3}: 19 of the 30 HVS in the northern
hemisphere seem to share a common travel time (see bin centered at
$\sim 133$ km/s). The peak in the histogram is higher than expected
from a model where ejections occur uniformly in time. The model
prediction is shown by the solid curve in the inset histogram, after
taking into account the volume surveyed, the finite lifetime of late
B-main sequence stars, and assuming the distribution of SMBH-ejection
velocities computed by Bromley et al (2006).

The significance of the peak seems high: selecting at random 30 stars
from such model yields a peak with 19 (or more) stars in fewer than 1
in 1000 trials.  The small number of objects involved precludes a
more conclusive assessment of this possibility, but we note that this
is not necessarily inconsistent with SMBH ejection. In this scenario,
a ``preferred travel time'' may just reflect a burst of star formation
that boosted the population of binaries in the Galactic center region
a few hundred Myr ago. We discuss a different interpretation below.

\begin{figure}
\epsscale{.89}
\begin{center}
\plotone{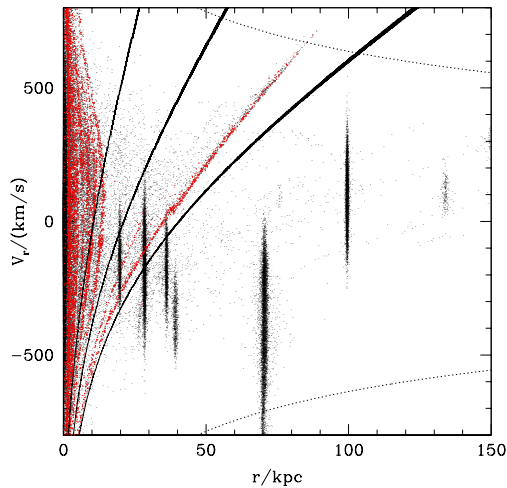}
\end{center}
\caption{
Galactocentric radial velocity vs radius for star particles in the
simulated galaxy of Abadi et al (2003a,b). Dots in red highlight stars
belonging to a satellite that merged recently with the main galaxy. As
in Fig.~\ref{figs:vr_r}, dotted lines show the escape velocity and
solid lines denote loci of constant travel time. Note how the tidal
debris stripped from the dwarf forms a stream of stars of nearly
constant travel time and with velocities which reach and exceed the
escape velocity from the system. This shows that hypervelocity stars
may also be produced during the disruption of a dwarf galaxy.
\label{figs:vallstr}}
\end{figure}

\begin{figure}
\epsscale{0.96}
\begin{center}
\plotone{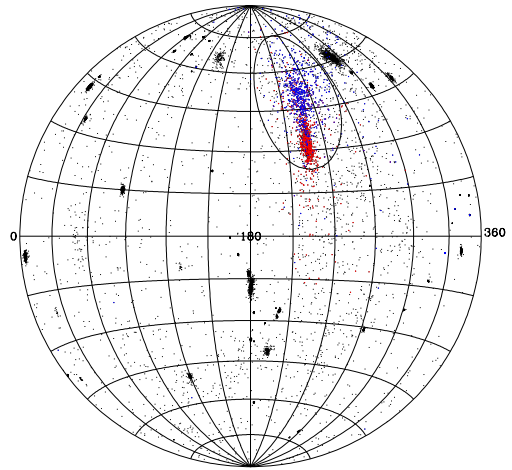}
\end{center}
\caption{
Aitoff projection of stars in Fig.~\ref{figs:vallstr}, after excluding
all stars in the main body of the galaxy (i.e., those with $r<20$
kpc), as seen by an observer at the Sun's location. Stars in red are
those belonging to the disrupted satellite; those in blue highlight
the HVS ($V_r>400$ km/s) component of the tidal stream. Note that the
tidal stream results in an overdensity of high-velocity stars
clustered tightly around the location of the tip of the stream and
that a gradient in the sky results from the projection of the stream
seen nearly ``end on''. Projected on the sky, the stream spans a few
tens of degrees and is roughly fully contained within a
$26^{\circ}$-radius circle such as that shown in
Fig.~\ref{figs:brown3}.
\label{figs:aitoff}}
\end{figure}

\subsection{HVS angular distribution}

Fig.~\ref{figs:brown3} shows an aitoff projection in Galactic
coordinates of all stars in the BGKK survey. High-speed stars
(including those reported recently by BGK08) are shown in color, coded
according to their travel times, as shown in the bottom-left
inset. The main point to note in this figure is that the HVS
population is not uniformly distributed in the sky, and that its
angular distribution appears inconsistent with a random sample of
stars in the survey.

One way of quantifying this is to focus on the region around the
constellation of Leo, which seems to contain the majority of HVS. We
choose for this a 26$^{\circ}$-radius circle centered at the position
of Leo I, itself a high-speed (possibly unbound) distant satellite
companion of the Milky Way (e.g. Sales et al. 2007a, and Mateo,
Olszewski \& Walker 2008). A total of $192$ stars in the BGKK survey
fall within the circle, or $21\%$ of the total of $935$ halo stars in
the northern section of the BGKK survey.  By contrast, $42\%$ of all
stars with $V_r>280$ km/s fall within the vicinity of Leo I, a
proportion that rises to $54\%$ for stars with $V_r>350$ km/s.

We may estimate the significance of this result by repeating the same
calculation after randomly reassigning the measured radial velocities
amongst the stars in the survey. A proportion at least as high as
$42\%$ of $V_r>280$ km/s stars falls within the circle in fewer than
$1$ in $100$ such random trials
\footnote{
Including the new high-velocity stars identified by BGK08 supports
this conclusion. The chance of observing such clustering around Leo I
goes down to roughly $1$ in $700$ for $V_r>280$ km/s stars, and
decreases even further if the center of the circle is moved from Leo I
in the direction of Leo IV. These estimates assume that the stars in
the BGK08 and BGKK parent surveys have similar distributions in the
sky.}
. For $V_r>350$ km/s stars, the corresponding number is fewer than $1$
in $125$ trials. Also, since the majority of HVS are at Galactic
longitude $l\sim 240^{\circ}$ (see Fig.~\ref{figs:brown3}) it is
unlikely that the HVS anisotropy is the result of the slightly larger
effective volume surveyed in the direction of the anti-Galactic
center. This supports the idea that the enhanced clustering of HVS in
the direction of Leo is real. The observed angular clustering is not
easily explained in the SMBH-ejection scenario, which predicts an HVS
population distributed approximately isotropically across the sky.

\subsection{A tidal debris origin for hypervelocity stars?}

The anisotropic distribution and the preferred travel time discussed
in the preceding subsections suggest that at least part of the HVS
population may have a different origin from the SMBH-ejection
mechanism. We explore here an alternative scenario where these two
peculiarities are reproduced naturally. Our proposal envisions
high-velocity stars in the halo as a result of the tidal disruption of
a dwarf galaxy in the Galactic potential.

This is illustrated in Fig.~\ref{figs:vallstr}, where we show the
position of ``star'' particles in a cosmological simulation of the
formation of a galaxy in the $\Lambda$CDM cosmogony. The simulation is
one in the series presented by Steinmetz and Navarro (2002) and
analyzed in detail in Abadi, Navarro \& Steinmetz (2006), where we
refer the interested reader to for further technical details.
Fig.~\ref{figs:vallstr} shows the position of all stars in the
$r$-$V_r$ plane, after rescaling the system to a virial velocity of
$V_{200}=220$ km/s. This scaling allows us to compare the results
directly with observations of the Milky Way. As in
Fig.~\ref{figs:vr_r}, the dotted lines indicate the escape velocity of
the system, and the solid curves indicate loci of constant travel
time.

The data in Fig.~\ref{figs:vallstr} are shown at a snapshot chosen a
couple of hundred Myr after the final merger/disruption of a dwarf
satellite in the potential of the main galaxy. The stars of the
disrupted satellite are shown in red in order to distinguish them
from the rest of the stars in the simulated galaxy, shown in
black. Note that a long stream of stars formerly belonging to the
satellite follow closely a line of constant travel time. These are
stars that were stripped from the dwarf during its last pericentric
passage, just before its main body merged with the central galaxy. The
travel-time coherence in the stream (or ``tidal tail'') results from
the fact that all those stars were stripped from the dwarf at the same
time. Stars along the stream have different energies, but it is clear
that the disruption event has been able to push some stars into very
high-velocity orbits, some even exceeding the local escape speed from
the system.

>From the perspective of an observer near the center of the galaxy
(such as an observer at the Sun in the Milky Way) the stream would be
seen projected onto a well-defined direction in the sky. This is shown
in Fig.~\ref{figs:aitoff}, which shows a projection on the sky of all
stars in the halo of the simulated galaxy (i.e., excluding stars with
distances less than $20$ kpc from the center). The points in color
correspond to the tidal debris from the disrupting dwarf. In
particular, blue dots highlight those with radial velocity exceeding
$400$ km/s. Note that, from the Sun's perspective, the ``tidal tail''
of high velocity stars stripped from the dwarf falls within a
well-defined region spanning a few tens of degrees in the sky.

Figs.~\ref{figs:vallstr} and ~\ref{figs:aitoff} thus illustrate that
tidal disruption of a dwarf galaxy is able to push stars to speeds
high enough to escape the galaxy, and provides a mechanism to populate
the halo with high-velocity stars with kinematic and angular
clustering peculiarities resembling those of the HVS population
discussed above. This suggests that at least some stars of the HVS
population may have originated in the recent disruption of a dwarf
galaxy.

How can this rather unorthodox proposal be validated/ruled out?
Fortunately, the most natural predictions of this scenario differ
significantly from those motivated by the SMBH-ejection mechanism. In
particular, a broader search for high-speed stars should confirm the
enhancement of HVS in the constellation of Leo. High-speed stars
should also exist amongst all spectral types, and should contain a
significant number of evolved stars and stars of low metallicity,
which should make up the bulk of the stellar content of the
dwarf. With enough statistics, a radial-velocity angular gradient
amongst high-speed stars might also be observed, such as that shown in
Fig.~\ref{figs:aitoff}.  The tail should also contain returning stars
with negative velocity that are relatively nearby (see, e.g.,
Fig.~\ref{figs:vallstr}) so it would be worth checking existing
surveys of nearby stars for unusual velocity patterns in the same
general direction in the sky.

We note that our argumentation rests on the assumption that the halo
of the Milky Way is relatively massive (i.e., $V_{200}\sim 220$ km/s).
This is suggested by semianalytic galaxy formation models and is
consistent with timing argument mass estimates for the Milky Way (Li
\& White 2007). There is evidence, however, that the virial velocity
of the Galactic halo might be considerably smaller ($V_{\rm vir} \la
140$ km/s), as indicated, for example, by recent analysis of Milky Way
satellite data (Sales et al 2007b), of RAVE data for solar
neighborhood stars (Smith et al 2007), and of SEGUE data on halo
stars (Xue et al 2008). This issue remains unresolved, but should a
low mass for the Milky Way halo be confirmed, it would mean that a
large fraction of HVS would be truly unbound and would argue against
the tidal debris interpretation proposed here.

Although examining further a tidal debris origin for HVS appears
worthwhile, it should be recognized that this proposal is not without
shortcomings. One of the main difficulties is to explain why a tidal
stream emanating from a dwarf should contain short-lived, massive
main-sequence stars, unless the accreting dwarf was gas-rich and
underwent a burst of star formation, perhaps triggered by pericentric
passages prior to its final disruption. Gas-rich dwarfs tend to be
rather massive, and careful modeling is needed in order to examine
whether the timescales really work out and to explain why the remains
of the putative dwarf have escaped detection so far. Systematic
radial-velocity surveys of large numbers of faint stars across vast
regions of the sky, such as those being planned by the GAIA satellite,
will undoubtedly be able to settle these questions in the foreseeable
future.

\acknowledgments
MGA acknowledges Laura V. Sales for a careful reading and for comments
on an early draft of this paper. We thank the anonymous referee for a
constructive report and acknowledge useful discussions with Warren
Brown during the KITP Conference ``Back to the Galaxy II''. This
research was supported in part by the National Science Foundation
under Grant No. PHY05-51164.


\end{document}